\documentclass[12pt]{JHEP}

\title{Wilson Loops as Matrix Strings} 
\author{Nadav Drukker \\ 
Institute for Theoretical Physics, University of California,\\
Santa Barbara, CA 93106\\
\smallskip
and\\
\smallskip
Department of Physics, Princeton University,\\
Princeton, NJ 08544\\
\email{drukker@itp.ucsb.edu}
}
\abstract{
In the framework of Matrix theory we show that Wilson loops can serve 
as interpolating fields to define string scattering amplitudes 
as gauge theory observables.
}
\preprint{NSF-ITP-99-125 \\ hep-th/9910211}

\newcommand{\beq}{\begin{equation}}
\newcommand{\eeq}[1]{\label{#1}\end{equation}}
\newcommand{\ber}{\begin{eqnarray}}
\newcommand{\eer}[1]{\label{#1}\end{eqnarray}}

\newcommand{\Tr}{\mbox{Tr}}
\newcommand{\cP}{{\cal P}}
\newcommand{\cS}{{\cal S}}
\newcommand{\half}{{1\over 2}}

\newcommand{\vev}[1]{{\left<#1\right>}}

\newfont{\Bbb}{msbm10 scaled 1200}     
\newcommand{\mathbb}[1]{\mbox{\Bbb #1}}

\begin{document}

\section{Introduction}

Matrix theory \cite{Banks:1997vh} is a proposed formulation of 
M-theory in the light cone gauge, or for finite $N$  the 
discrete light cone quantization of M theory 
\cite{Susskind:1997cw}. When one of the target space 
directions is compactified it turns into a non perturbative 
formulation of light cone type IIA string theory 
\cite{Motl:1997th,Banks:1997my,Dijkgraaf:1997vv}. 
In the UV this theory is two 
dimensional supersymmetric $U(N)$ gauge theory, the dimensional 
reduction of ten dimensional SYM.

The theory lives on a cylinder parameterized by $\sigma,\tau$ where 
$\sigma$ has period $2\pi$. 
There are two gauge fields $A^\alpha$, eight scalars 
$X^i$ ($i=2,\ldots 9$) and their superpartners $\Psi$. The action is
\ber
\cS&=&
\int d\tau d\sigma\,\Tr\left[
{1\over 4g_{YM}^2} F_{\alpha\beta}^2
+\half\left(D_\alpha X^i\right)^2
+{g_{YM}^2\over4}[X^i,X^j]^2
\right.
\nonumber\\
&&~~~~~~~~~~~~~~~~~~\left.
+i\bar\Psi\Gamma^\alpha D_\alpha\Psi
+ig_{YM}\bar\Psi\Gamma^i[X^i,\Psi]\right].
\eer{action}

This action is a non-Abelian generalization of the light cone 
Green-Schwarz string action. In the infra-red $g_{YM}\to\infty$, 
the potential energy 
$[X^i,\,X^j]^2$ is minimized by commuting matrices, so the action 
reduces to a collection of free strings. The only subtlety is that 
the holonomy around the cylinder does not have to be trivial. When 
this happens, instead of $N$ ``short'' strings one gets fewer 
``long'' strings. In the large $N$ limit, those long strings 
become the usual light cone strings.

It was shown \cite{Dijkgraaf:1997vv} that the least irrelevant 
operator with all the necessary symmetry that can be added to this 
theory corresponds to a joining/splitting interaction of the strings.

In \cite{Giddings:1998yd} those interactions were further studied. It 
was shown that the singular points in light cone string diagrams are 
smoothed by a non-commutative region. At high energy the string 
scattering amplitude is dominated by the Gross-Mende surfaces 
\cite{Gross:1987kz}. For those surfaces the interaction point is 
replaced by an instanton solution of the self-dual YM equations.

In this approach one solves the (self-dual) YM equations for given 
boundary conditions.  The solutions are instantonic configurations 
carrying a charge in the permutation group~$S_N$ relating the incoming 
arrangement of the string bits to the outgoing one.  For gauge 
invariance it is necessary to sum over all permutations of the $N$ 
components of the incoming and outgoing strings.  Part of this sum, 
say over the the arrangement of the incoming strings is trivial, but 
a large part of the sum is non trivial.  There will be many distinct 
configurations (as many as $\sim N!$) which correspond to the same 
physical process.  Each of them is related to a different instanton.

To calculate the scattering amplitude in this way will require a sum 
over a very large number of different processes.  It is possible that 
the sum is dominated by a small number of saddle points, but we don't 
know that.  Otherwise, this seems like a very difficult problem, 
perhaps the sum does not converge.

In this paper we try a different approach.  Instead of solving the 
classical equations of motion, whose solutions are not gauge 
invariant, we find gauge invariant operators which are the non-Abelian 
generalization of string vertex operators.  Using those operators, 
which are certain Wilson loops, we can define a string scattering 
amplitude as a gauge theory observable. These 
correlators can be regarded as the non-perturbative string scattering 
amplitude, and are well defined objects in the gauge theory.  
Unfortunately, we don't know how to evaluate them, except by
reducing back to a calculation like \cite{Giddings:1998yd}.

\section{String vertex operators}\label{vertex-section}

We will be studying the Wilson loop operator
\beq
W(p_i,p_-)={1\over p_-}\Tr\,{\cal P}
\exp\left(
i\oint_C\left(A_1+{p_i\over p_-}X^i\right)d\sigma
\right).
\eeq{vertex1}
The curve $C$ wraps the world sheet $p_-$ times along a straight line 
in the $\sigma$ direction.  $p_i$ are arbitrary parameters that 
couple to the scalars in the loop (they could be functions of $\sigma$).

To see what happens to the loop in the IR we should evaluate it 
for the fields that remain in that limit.
The configurations that survive are the long strings, they 
have a gauge field $A_\sigma(\sigma^\alpha)$ for which 
$\exp(i\int_0^{2\pi}A_\sigma(\sigma^\alpha)\,d\sigma^1)=U$ is in 
the Weyl group of $U(N)$ (a permutation matrix).  At every point all 
the~$X^i$ commute, and when parallel transported, also at different 
points.  It is possible to make a large gauge transformation that will 
set~$A_\sigma$ to zero and will make all the~$X^i$'s diagonal, but not 
single valued.  One has to be careful when doing that, though, since 
the operator~(\ref{vertex1}) is invariant only under small gauge 
transformations and not under large ones.  Instead one is left 
with~$U^{p_-}$ in the trace %
\beq
W(p_i,p_-)\rightarrow{1\over p_-}\Tr\left[
\exp\left(i\int_0^{2\pi p_-}
{p_i\over p_-}\tilde X^i(\sigma)\,d\sigma\right)
U^{p_-}\right],
\eeq{vertex2}
where~$\tilde X^i$ is~$X^i$ after it was diagonalized, and 
it's defined for~$\sigma$ beyond~$2\pi$ through the twisted boundary 
conditions.

If $U$ includes a cycle of length~$p_-$, then in that block~$U^{p_-}$ 
is the unit matrix $I_{p_-}$ and~$\tilde X^i$ are (after reordering)
\beq
\tilde X^i(\sigma)=\pmatrix{
x^i(\sigma) & 0                & \cdots & 0                         \cr
0           & x^i(2\pi+\sigma) & \cdots & \vdots                    \cr
\vdots      &                  & \ddots & \vdots                    \cr
0           & \cdots           & \cdots & x^i(2\pi(p_--1)+\sigma) }.
\eeq{tilde-X}
It's clear that~$\int_0^{2\pi p_-}\tilde X^i(\sigma)$ in that block 
is proportional to the identity matrix, so the trace of the exponent 
is just $p_-$ times the exponent. Therefore on this block the Wilson 
loop reduces to
\beq
W(p_i,p_-)\rightarrow
\exp\left(i \int_0^{2\pi p_-}
{p_i\over p_-}x^i(\sigma)\,d\sigma\right).
\eeq{vertex3}
This is precisely the form of the vertex operator for the light-cone 
string \cite{Mandelstam:1985ww}! If $p_i$ are constants, this is the 
vertex operator for the graviton.

There is a subtle point when~$U$ contains a cycle of 
length~$n$ which divides~$p_-$. Clearly when~$U$ contains no such 
cycles, the Wilson loop vanishes as expected. But as defined 
above, it will not give zero for configurations with cycles of length 
that divides~$p_-$.

To eliminate that problem one could recursively define a modified 
Wilson loop operator with the extra terms subtracted. Alternatively one 
could approximate~$p_-$ by a prime number, which 
should be possible for large enough matrices.  That only leaves the 
case of cycles of length~1.

Instead, we think that this effect should be regarded as a real 
artifact of the finite $N$ theory. This is consistent 
with the observation \cite{Giddings:1998yd} that the matrix 
string graviton seems to include both 
a long string and many short strings.  It also fits well with 
what's known on the nature of the wave function for bound states of 
$N$ short strings \cite{Green:1998tn}. There too, for non prime $N$ 
one gets contributions from all the divisors of $N$.

We just showed that the Wilson-loop (\ref{vertex1}) turns into the 
corresponding string vertex operator in the infra-red. Since the 
YM action reduces to the string action the correlation function
\beq
\vev{W_1\cdots W_n},
\eeq{corr}
with some of the operators inserted at negative infinite time, and 
some at positive infinite time is the string 
scattering amplitude in Matrix theory. This is the non-perturbative 
definition for the scattering amplitude of strings with 
corresponding momenta.

One should note that the Wilson loops $W$ are 
{\em not} the wave function of the matrix string.  In particular they 
have large support on non-commuting configurations.  Likewise, those 
operators are not the only operators one could use.  Adding to $W$ any 
function whose support does not include the ``long strings'' 
configurations will not alter this result.  Any such modification 
would work perfectly well as a vertex operator, the difference 
corresponds to more massive modes that will not modify the calculation 
when the incoming and outgoing states are sufficiently far apart.

\section{Discussion}

We showed that the Wilson loop operators (\ref{vertex1}) can serve 
as interpolating fields for strings. Using them we defined the Matrix 
string scattering amplitude as a gauge invariant observable (\ref{corr}). 
One would like to find a way to evaluate those correlators.

This turns out to be a difficult problem, those Wilson loops are not 
simple operators. If one tries to use perturbation theory, it 
is natural to expand about classical solutions---the long strings 
configurations. This reduces the problem back to the calculation 
of \cite{Giddings:1998yd}. One has to sum over many ingoing and outgoing 
classical solutions, and calculate the instantonic processes that take 
one to the other.

We have not found any other way to calculate these correlators.

Apart from strings, matrix theory contains other objects 
\cite{Banks:1997nn}. It is easy to write down operators which have the 
appropriate charges. For example, the Wilson loop
\beq
W=\Tr\,\cP\oint_Cd\sigma\, F_{01}
\exp\left(i\oint_C\left(A_1+{p_i\over p_-}X^i\right)d\sigma\right),
\eeq{0-brane}
can be used as a vertex operator for a 0-brane.

Likewise
\beq
W=\Tr\,\cP\oint_C d\sigma [X^i,X^j]
\exp\left(i\oint_C\left(A_1+{p_i\over p_-}X^i\right)d\sigma\right),
\eeq{2-brane}
is a Wilson-loop for a 2-brane in the~$i,j$ plane.

\acknowledgments
I wish to thank David Gross for a lot of help. 
This work was supported in part by the NSF under grant No. PHY94-07194.

\end{document}